# Design of an efficient mid-wave IR fiber optic light source

**A. Barh, S. Ghosh, R. K. Varshney, and B. P. Pal[*]**
*Department of Physics, Indian Institute of Technology Delhi, New Delhi, India – 110016.*
*\*Corresponding author: bppal@physics.iitd.ac.in*

**Abstract:** Design of a mid-wave IR broadband (3.1 – 3.8 μm) fiber-based light source exploiting FWM in a meter long suitably designed chalcogenide microstructured fiber is reported. An achievable gain more than 35 dB is demonstrated.
**OCIS codes:** (060.4370) Nonlinear optics, fibers; (190.4380) Nonlinear optics, four-wave mixing; (060.4005) Microstructured fibers; (160.4330) Nonlinear optical materials.

## 1. Introduction

In recent years, there has been a strong interest to leverage on huge development made in fiber optic telecommunication and extend that knowledge to develop fibers and fiber-based devices suitable for mid-IR spectral region (2-10 μm). Emerging potential applications like non-destructive soft tissue ablation in medical diagnostics, monitoring of combustion flow and gas dynamics through molecular absorption spectroscopy, semiconductor processing (e.g. in-situ real time monitoring of plasma etch rates), and huge military applications in the mid-wave IR (MWIR) spanning 3-5 μm region have lately attracted a lot of research investments. MWIR wavelength region is particularly important since a large number of molecules undergo strong characteristic vibration band transitions in this domain, which is also known as "molecular fingerprint regime" e.g. various hydrocarbons, hydrochlorides and commonly used solvents show strong absorption in the range of 3.2 – 3.6 μm [1]. Compact fiber-based light sources for MWIR would find wide scale military applications as it is a clean atmospheric window for high power transmission leading to applications in heat sinking missiles, IR counter-measures, and also low power applications in night vision.

In this paper, our aim is to numerically design a broad-band (covering 3-4 μm spectral domain) and high gain mid-IR source by exploiting the four-wave mixing (FWM) process through the extraordinary linear and nonlinear (NL) properties of chalcogenide glass (S-Se-Te)-based microstructured optical fibers (MOF), for which various well matured fabrication technologies exist though expensive in practice [2]. Through appropriate dispersion tailoring, the FWM band width (BW) and over-all gain in the fiber can be maximized; thereby a broad-band fiber-based light source for a targeted wavelength regime like MWIR can be designed for potential fabrication.

## 2. Design methodology

Under the FWM process, two pump photons of wavelength $\lambda_P$ get converted/shifted into a signal photon ($\lambda_s$) and an idler photon ($\lambda_i$). For efficient mixing, the phase matching condition is very important, and the fiber's dispersion profile should be so designed that it's zero dispersion wavelength ($\lambda_{ZD}$) falls within the emitting wavelength(s) of available high power light source(s). Under continuous wave (CW) pump condition, considering up to 4$^{th}$ order dispersion term ($\beta_4$), the maximum frequency shift ($\Omega_s$) can be expressed as [3]

$$\Omega_S^2 = \frac{6\beta_2}{\beta_4}\left[-1 \pm \sqrt{1 - \frac{2\beta_4 \gamma P_0}{3\beta_2^2}}\right] \quad (1)$$

where, $\beta_m$ is the $m^{th}$ order group velocity dispersion term, $\gamma$ is the well-known effective NL coefficient, and $P_0$ is average pump power. For given values of $\gamma$ and $P_0$, $\Omega_s$ and hence, the BW can be optimized by choosing appropriate combination of $\beta_2$ and $\beta_4$. Specifically for continuous broad-band gain on either side of pump wavelength, the pump must experience low anomalous dispersion ($\beta_2 \leq 0$). In addition, to maintain the phase matching condition over a broad wavelength range, positive $\beta_4$ would be used to counter the phase matching induced by negative $\beta_2$. Thus higher order dispersion management is very crucial in such fiber designs; positive $\beta_4$ leads to broad-band and flat gain where as negative $\beta_4$ reduces the flatness and BW of FWM [4]. This positive $\beta_4$ value in the vicinity of low negative $\beta_2$ could be achieved by suitably reducing the core cladding index difference ($\Delta n$). Hence through a proper choice of the materials for MOF, multi-order dispersion management is feasible to engineer the FWM efficiency.

To achieve such tailored application-specific fiber designs, we focus on an arsenic sulphide ($As_2S_3$)-based MOF geometry with a solid core and holey cladding, consisting of 4 rings of hexagonally arranged holes embedded in $As_2S_3$ matrix. In order to reduce $\Delta n$, we assume that borosilicate glass rod would fill the holes. Compatible thermal properties of the $As_2S_3$ and borosilicate glass should allow feasibility of fabrication of such a holey MOF [5]. The wavelength dependence of the linear refractive index of $As_2S_3$ and borosilicate glass has been incorporated through Sellmeier formula [5]. To attain sufficient signal gain ($G_s$) over the MWIR regime of 3.1 ~ 3.8 μm, we have assumed commercially available CW $Er^{3+}$-doped ZBLAN fiber laser emitting at 2.80 μm [1] as the pump and confined our numerical study to pump power levels below 5 W to suppress other potential NL effects [3, 4].

Using above-mentioned fiber design methodology, we have optimized the MOF with fiber parameters as $d/\Lambda$ = 0.5 and $\Lambda$ = 2.5 μm. Spectral response of GVD parameters $\beta_2$ and $\beta_4$ are shown in Figs. 1(a) and (b), respectively. Figure 1 clearly indicates that $\lambda_{ZD}$ falls at 2.792 μm. At a fixed $P_0$, maximum gain ($G_{s,max}$) depends on $\gamma$ and length of the fiber ($L$) as $G_s$ increases exponentially with "$\gamma P_0 L$". But BW is inversely proportional to this $L$ as long as $L >> L_{NL}$ (= $(\gamma P_0)^{-1}$). Thus to maintain large BW, $L$ should be relatively short at the cost of peak gain.

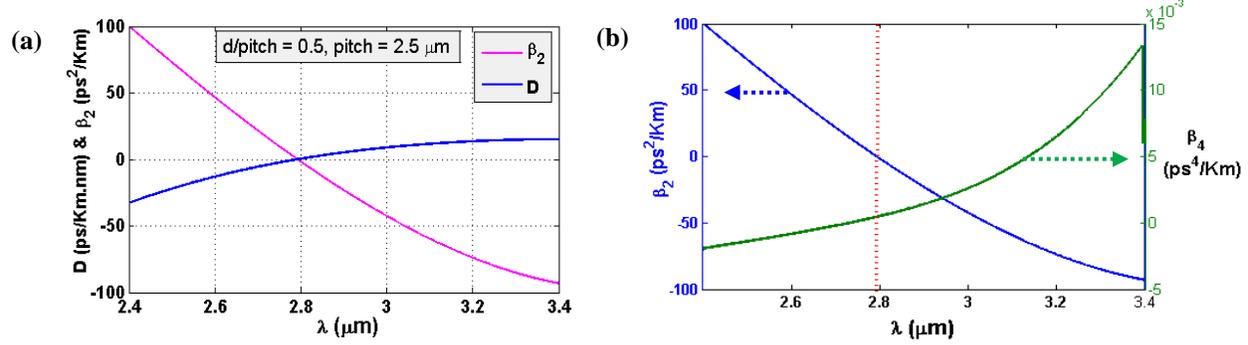

Fig. 1. Dispersion characteristics of $As_2S_3$ and Borosilicate based solid core MOF for $d/\Lambda$ = 0.5 and $\Lambda$ = 2.5 μm. (a) Group velocity dispersion D (blue curve) and $\beta_2$ (pink curve) variation with operating wavelength ($\lambda$); $\lambda_{ZD}$ = 2.792 μm. (b) Variation of $\beta_2$ (blue curve) and $\beta_4$ (green curve) with $\lambda$. Red dotted line indicates the desired pump wavelength.

## 3. Results and discussions

During optimization of high-flat gain and maximum BW, a strong interplay was evident amongst $P_0$, $L$, and $\lambda_P$. If we detune $\lambda_P$ from $\lambda_{ZD}$, absolute value of $\beta_2$ increases leading to fluctuations in the gain spectrum due to change in $\Delta\kappa$ around its zero value. This makes the spectrum more discrete as shown in Fig. 2; where only upper half (signal side) of gain spectrum is displayed. In the vicinity of $\lambda_P$, the gain decreases as $\lambda_s$ approaches $\lambda_P$. The phase mismatch term becomes $2\gamma P_0$ and the $G_s$ becomes $(1+ \gamma P_0 L)$, which lead to the linear growth of signal from $\lambda_P$. From this figure, it can be interpreted that the best result could be achieved for $\lambda_P \approx 2.797$ μm, where BW can be maximized but have to take care of its flatness and $G_{s,max}$. Calculated $A_{eff}$ at this wavelength is ~ 9.205 μm².

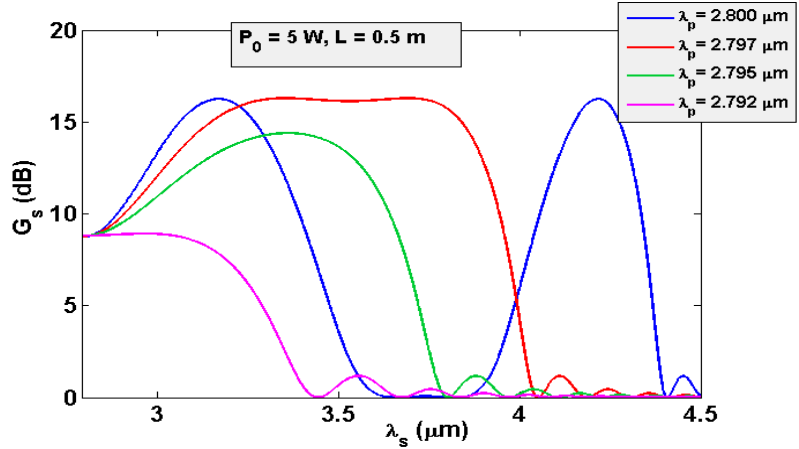

Fig. 2. Variation of signal gain for different pump wavelengths. With pumping at $\lambda_{ZD}$ (= 2.792 μm), gain spectrum is almost uniform around $\lambda_P$. With increase in $\lambda_P$ from $\lambda_{ZD}$, the BW as well as fluctuation increases.

Using 2.797 μm as the pump, and fixing $P_0$ at 5 W, we have studied the variation of gain spectrum for different values of fiber length (shown in Fig. 3(a)). This figure clearly indicates that the maximum gain increases with increase in $L$ but at the cost of narrower BW. Thus to obtain high gain of more than 35 dB with full width at half maxima (FWHM) of ~ 670 nm, optimum set of parameter were found to be $P_0$ = 5 W, $L$ = 1 m, and $\lambda_P$ = 2.797 μm (shown in Fig. 3(b)). The confinement loss over the entire BW is very small (< 0.01 dB/m).

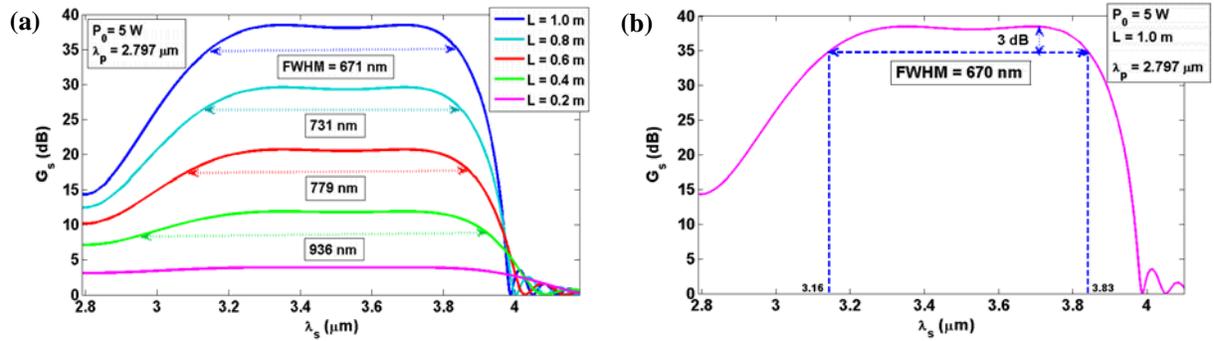

Fig. 3. Gain spectrum for generated signal with pumping at 2.797 μm. (a) Variation of $G_s$ for different $L$ (0.2 – 1.0 m) with fixed $P_0$ at 5W. (b) Gain spectrum for optimum parameter. Achieved maximum gain is 38.5 dB and FWHM is almost 670 nm.

Further, for attaining broader BW while maintaining $P_0$ below 5W, we have to include the idler wavelength side as well to calculate the gain spectrum. The entire gain spectrum for different $\lambda_p$ is shown in Fig. 4(a). The most important non-uniformity in such gain spectrum is "the dip in the vicinity of $\lambda_P$" and we have to minimize the difference between $G_{s,max}$ and gain at pump ($G_p$) to reduce this non-uniformity. As with the increment of $P_0$, this difference ($\Delta G$) increases, we confined our calculation to $P_0 \sim 3$W. From this figure, it can be interpreted that best results could be achieved for $\lambda_P \approx 2.796$ μm (red curve), where BW can be maximized; however one needs to take in to account its flatness. Variation of gain spectrum for different $P_0$ at $\lambda_P \approx 2.796$ μm was studied and found that for $P_0 = 3$W, the FWHM is maximum and the $\Delta G$ is less than 3 dB. Further optimization of $L$, yields the optimum set of parameters as $P_0 = 3$W, $L = 50$ cm, for which the $G_{s,max} \approx 7.72$ dB and $\Delta G \sim 2.4$ dB were achieved (see Fig. 4(b)). In this case, FWHM can be achieved as high as 1300 nm ranging from 2.3 μm to 3.6 μm with an average gain > 5 dB.

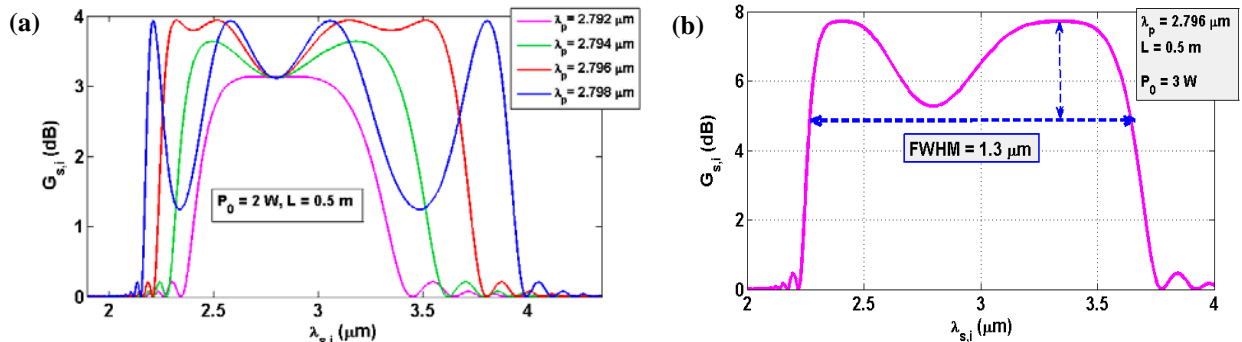

Fig. 4. (a) Variation of $G_{s,i}$ for different pump wavelengths. For $P_0 = 2$W and $L = 50$ cm. (b) Gain spectrum for optimum parameters. Achieved maximum gain was 7.72 dB and $\Delta G$ was ~ 2.4 dB.

In conclusion, in this paper, we report a theoretical design of a broad-band mid-IR light source by maximizing FWM band-width and gain in a highly nonlinear chalcogenide MOF. For high power broad-band source, a source of 3.1 – 3.8 μm with gain more than 35 dB can be generated for a 2.797 μm pump of 5W average power in 1 meter long sample of the designed fiber. This designed fiber should be a good candidate for MWIR applications.

This work relates to Department of the Navy Grant N62909-10-1-7141 issued by Office of Naval Research Global. The United States Government has royalty-free license throughout the world in all copyrightable material contained herein. One of the authors A.B. gratefully acknowledges award of Ph.D. fellowship by CSIR (India).